\documentclass[12pt,A4]{article}
\usepackage{amssymb,amsfonts,amsthm,amsmath}

\usepackage{epsf}
\usepackage{graphicx}
\usepackage{placeins}
\usepackage{setspace}

\def\hang{\hangindent\parindent}
 \def\rf{\par\noindent\hang}

\begin{document}

\baselineskip=21pt

\begin{center} \Large{{\bf
Simultaneous Confidence Intervals for the Population Cell Means, for Two-by-Two
Factorial Data, that Utilize Uncertain Prior Information}}
\end{center}

\bigskip

Short running title: Confidence Intervals from Two-by-Two Factorial Data

\bigskip


\begin{center}
\large{ PAUL KABAILA$^{1*}$ {\normalsize \rm AND} KHAGESWOR GIRI$^2$}
\end{center}

\bigskip
\bigskip

\begin{center}
{$^1$Department of
Mathematics and Statistics, La Trobe University, Victoria,
Australia}
\end{center}

\smallskip

\begin{center}
{$^2$
Department of Primary Industries,
600 Sneydes Road, Werribee, Victoria, Australia}
\end{center}

\bigskip

\vbox{\vskip 10cm}


\noindent $^*$ Author to whom correspondence should be addressed.

\medskip

\noindent Department of
Mathematics and Statistics, La Trobe University, Victoria 3086,
Australia; tel.: +61-3-9479-2594; fax: +61-3-9479-2466;
E-mail: P.Kabaila@latrobe.edu.au.

\newpage

\begin{center}
{\bf Abstract}
\end{center}

\medskip

\noindent Consider a two-by-two factorial experiment with more than 1 replicate.
Suppose that we have uncertain prior information that the two-factor interaction
is zero. We describe new simultaneous frequentist confidence intervals for the 4 population cell
means, with simultaneous
confidence coefficient $1-\alpha$, that utilize this prior information in the following sense.
These simultaneous confidence intervals define a cube with expected volume that (a) is
relatively small when the two-factor interaction is zero and (b) has maximum value that is not
too large. Also, these intervals coincide with the standard simultaneous confidence intervals obtained
by Tukey's method, with simultaneous
confidence coefficient $1-\alpha$, when the data strongly contradict the prior information
that the two-factor interaction is zero. We illustrate the application of these new
simultaneous confidence intervals to a real data set.

\bigskip
\medskip


\newpage

\noindent {\bf 1. Introduction}

\medskip

\noindent Hodges and Lehmann (1952), Bickel (1984) and Kempthorne (1983, 1987, 1988) present frameworks
for the utilization of uncertain prior information (about the parameters of the model)
in frequentist inference, mostly for point estimation.
We say that the confidence set ${\cal C}$ is a $1-\alpha$ confidence
set for the parameter of interest
if its infimum coverage probability is
$1-\alpha$. We assess such a confidence
set by its scaled expected volume, defined to be
(expected volume of ${\cal C}$)/(expected
volume of the standard $1-\alpha$ confidence set).
The first requirement of a
$1-\alpha$ confidence set that utilizes the uncertain
prior information is that its scaled expected volume is significantly less than 1 when the
prior information is correct (Kabaila, 2009).

Confidence sets that satisfy this first requirement can be classified
into the following two groups. The first group consists of
$1-\alpha$ confidence sets with scaled expected volume
that is less than or equal to 1 for all parameter values, so that these dominate the
standard $1-\alpha$ confidence set. Examples of such confidence sets are the Stein-type confidence interval
for the normal variance (see e.g. Maata and Casella, 1990 for a review) and Stein-type confidence sets for the multivariate
normal mean (see e.g. Saleh, 2006 for a review). The second group consists of $1-\alpha$
confidence sets that satisfy this first requirement, when dominance of the usual $1-\alpha$ confidence set is not possible (the
scaled expected volume must exceed 1 for some parameter values). This second
group includes confidence intervals described by Pratt (1961), Brown et al (1995)
and Puza and O'Neill (2006ab). This second group also includes
$1-\alpha$ confidence sets that satisfy the additional requirements that (a) the maximum (over the parameter space)
of the scaled expected volume is not too much larger than 1 and (b) the confidence set reverts to
the usual $1-\alpha$ confidence set when the data happen to strongly contradict the prior information.
Confidence intervals that utilize uncertain prior information and satisfy these additional requirements
have been proposed by Farchione and Kabaila (2008)
and Kabaila and Giri (2009a).

Consider a $2\times 2$ factorial experiment with $c$ replicates, where $c>1$.
Label the factors A and B. Suppose that the parameters of interest are the
four population cell means $\theta_{00}, \theta_{10}, \theta_{01}, \theta_{11}$
where, for example, $\theta_{10}$ denotes the
expected response when factor A is high and factor B is low.
Also suppose that, on the basis of previous experience with similar data sets
and/or expert opinion and scientific background, we have uncertain prior information
that the two-factor interaction is zero. Our aim is to find
simultaneous frequentist confidence intervals for the population cell
means, with simultaneous
confidence coefficient $1-\alpha$, that utilize this prior information.

Throughout this paper, we find that the simultaneous confidence intervals
of interest define a cube. For convenience,
we henceforth refer to this cube, rather than the corresponding simultaneous
confidence intervals. Let $\theta = (\theta_{00}, \theta_{10}, \theta_{01}, \theta_{11})$.
The standard $1-\alpha$ confidence cube
for $\theta$
is found using Tukey's method (described e.g. on p.289 of Bickel and Doksum (1977)).
We assess a $1-\alpha$ confidence cube for
$\theta$
using the scaled expected volume of this confidence cube, defined to be the
ratio (expected volume of this confidence cube)/(expected volume
of standard $1-\alpha$ confidence cube). We say that this confidence cube
utilizes the prior information if it has the following desirable properties.
This confidence cube has scaled expected volume that
(a) is significantly less than 1 when the two-factor interaction is zero and (b) has a maximum
value that is not too much larger than 1. Also, this confidence cube coincides with the standard
$1-\alpha$ confidence cube when the data strongly contradict the prior information
that the two-factor interaction is zero.

 The development of this new
$1-\alpha$ confidence cube parallels the development by Kabaila and Giri (2009a) of a new
$1-\alpha$ confidence interval for a specified {\sl simple} effect, for $2 \times 2$ factorial
data, that utilizes uncertain prior information that the two-factor interaction is zero.
It is fortunate that the symmetries in the more complicated context considered in the present paper
lead to simplifications that make the computation of the new
$1-\alpha$ confidence cube feasible.

In Section 3 we provide
a numerical illustration of the properties of this new confidence cube for $1-\alpha = 0.95$
and $c = 2$. The two-factor interaction is described by the parameter $\beta_{12}$
in the regression model used for the experiment. The uncertain prior information
is that $\beta_{12} = 0$. Define the parameter
$\gamma = \beta_{12}/\sqrt{\text{var}(\hat \beta_{12})}$, where $\hat \beta_{12}$
denotes the least squares estimator of $\beta_{12}$.
The scaled expected volume of the
new confidence cube for $\theta$ is an even function
of $\gamma$. The bottom panel of Figure 2 is a plot of the square root of the
scaled expected volume of the new 0.95 confidence cube for $\theta$,
as a function of $\gamma$. When the prior information is correct (i.e. $\gamma=0$),
we gain since the square root of the scaled expected volume
is substantially smaller than 1. The maximum value of the square root
of the scaled expected volume is not too large.
The new 0.95 confidence cube for $\theta$
coincides with the standard 0.95 confidence cube
when the data strongly contradicts the prior information. This is reflected in Figure 2 by the fact that the
square root of the scaled expected volume approaches 1 as $\gamma \rightarrow \infty$.
In Section 4 we illustrate the application of the new $1-\alpha$ confidence cube to a
real data set.



\bigskip

\noindent{\bf 2. The Standard $\boldsymbol{1-\alpha}$ Confidence Cube Based on Tukey's Method}

\medskip

\noindent Let $Y$ denote the response and $x_1$ and $x_2$ denote the coded levels
 for factor A and factor B respectively, where $x_1$  takes values $-1$ and $1$ when the factor
A takes the values low and high respectively and $x_2$  takes
values $-1$ and $1$ when the factor B takes the values low and
high respectively. We assume the model
\begin{equation}
\label{regression}
Y = \beta_0 + \beta_1 x_1 + \beta_2 x_2 + \beta_{12} x_1 x_2 +
\varepsilon
\end{equation}
where $\beta_0$, $\beta_1$, $\beta_2$ and $\beta_{12}$ are
unknown parameters and the $\varepsilon$ for different response
measurements are independent and identically $N(0, \sigma^2)$
distributed. Because we are considering $c$ replicates, the number
of measurements of the response is $n=4c$. The dimension of the
regression parameter vector $(\beta_0, \beta_1, \beta_2, \beta_{12})$
is $p=4$. The parameters of interest are
$\theta_{00} = \beta_{0}-\beta_{1}-\beta_{2}+\beta_{12}$,
$\theta_{10} = \beta_{0}+\beta_{1}-\beta_{2}-\beta_{12}$,
$\theta_{01} = \beta_{0}-\beta_{1}+\beta_{2}-\beta_{12}$ and
$\theta_{11} = \beta_{0}+\beta_{1}+\beta_{2}+\beta_{12}$.

Let  $\big(\hat\beta_0$, $\hat\beta_1$, $\hat\beta_2,
\hat\beta_{12}\big)$ denote the least squares estimator of
$\big(\beta_0, \beta_1, \beta_2, \beta_{12}\big)$. Also, let
$\big(\hat \Theta_{00}, \hat \Theta_{10}, \hat \Theta_{01}, \hat
\Theta_{11} \big)$ denote  the least squares estimator of
$\big(\theta_{00}, \theta_{10}, \theta_{01}, \theta_{11}\big)$.
Note that $\big(\hat \Theta_{00}, \hat \Theta_{10}, \hat
\Theta_{01}, \hat \Theta_{11}, \hat\beta_{12} \big)$ has a
multivariate normal distribution with mean \newline
$\big(\theta_{00},
\theta_{10}, \theta_{01}, \theta_{11}, \beta_{12} \big)$ and
covariance matrix $\sigma^2 V$, where
\begin{equation*}
 V=\frac{1}{4c}\left[\begin{matrix}
  4 & 0 & 0 & 0 & 1 \\
  0 & 4 & 0 & 0 & -1 \\
  0 & 0 & 4 & 0 & -1 \\
  0 & 0 & 0 & 4 & 1 \\
  1 & -1 & -1 & 1 & 1 \\
\end{matrix}\right]
\end{equation*}
Let $v_{ij}$ denote the $(i,j)$ th element of $V$.
Also, let $\hat\sigma^{2}$ denote the usual estimator of $\sigma^{2}$
obtained by fitting the the full model to the data.
Define $W = \hat \sigma / \sigma$. Note that $W$ has the same distribution
as $\sqrt{Q/(n-p)}$ where $Q \sim \chi^2_{n-p}$.
Define $d_{1-\alpha}$ by
$P\big(\max_{i=1,\ldots, 4} |Z_i|/W \le d_{1-\alpha} \big ) = 1-\alpha$, where
$Z_1$, $Z_2$, $Z_3$, $Z_4$ and $W$ independent random
 variables and $Z_i \sim N(0,1)$ $(i=1,2,3,4)$.
 We use $[a \pm b]$ to denote the interval
 $[a-b, a+b]$. The standard
$1-\alpha$ confidence cube  for $\theta$, based on Tukey's method, is
\begin{multline*}
I = \left [ \hat \Theta_{00} \pm d_{1-\alpha} \sqrt{v_{11}}
\hat \sigma\right ]
\times\left [ \hat \Theta_{10} \pm
d_{1-\alpha} \sqrt{v_{22}} \hat \sigma\right ]\\
\times\left [
\hat \Theta_{01} \pm d_{1-\alpha} \sqrt{v_{33}} \hat
\sigma\right ]
\times\left [ \hat \Theta _{11}\pm d_{1-\alpha}
\sqrt{v_{44}} \hat \sigma\right ].
\end{multline*}

\bigskip

\noindent{\bf 3. New $\boldsymbol{1-\alpha}$ Confidence Cube that Utilizes the Prior
Information}

\medskip

\noindent The uncertain prior information is that $\beta_{12}=0$.
Define the confidence cube $J(b,s)$ for
$\theta$
to be
\begin{align}
\label{J(b,s)} &\left [ \hat \Theta_{00} - \sqrt{v_{11}} \hat
\sigma \, b\bigg(\frac{\hat\beta_{12}}{\hat \sigma
\sqrt{v_{55}}}\bigg) \, \pm \, \sqrt{v_{11}} \hat \sigma \,
s\bigg(\frac{|\hat\beta_{12}|}{\hat \sigma \sqrt{v_{55}}}\bigg)
\right ] \notag
\\
&\phantom{123}\times \left [ \hat \Theta_{10} +\sqrt{v_{22}} \hat
\sigma \, b\bigg(\frac{\hat\beta_{12}}{\hat \sigma
\sqrt{v_{55}}}\bigg) \, \pm \, \sqrt{v_{22}} \hat \sigma \,
s\bigg(\frac{|\hat\beta_{12}|}{\hat \sigma \sqrt{v_{55}}}\bigg)
\right ] \notag
\\
&\phantom{123456}\times \left [ \hat \Theta_{01} + \sqrt{v_{33}}
\hat \sigma \, b\bigg(\frac{\hat\beta_{12}}{\hat \sigma
\sqrt{v_{55}}}\bigg) \, \pm \, \sqrt{v_{33}} \hat \sigma \,
s\bigg(\frac{|\hat\beta_{12}|}{\hat \sigma \sqrt{v_{55}}}\bigg)
\right ] \notag
\\
&\phantom{123456789}\times \left [ \hat \Theta_{11} -
\sqrt{v_{44}} \hat \sigma \, b\bigg(\frac{\hat\beta_{12}}{\hat
\sigma \sqrt{v_{55}}}\bigg) \, \pm \, \sqrt{v_{44}} \hat \sigma \,
s\bigg(\frac{|\hat\beta_{12}|}{\hat \sigma \sqrt{v_{55}}}\bigg)
\right ],
\end{align}
where the functions $b$ and $s$ are required to satisfy the
following restriction.
\smallskip

\noindent {\underbar{Restriction 1}} \ \
$b: \mathbb{R} \rightarrow \mathbb{R}$  is
an odd function and $s: [0, \infty) \rightarrow [0, \infty)$.

\noindent Giri (2008) provides invariance arguments, of the type used by Farchione and Kabaila (2008)
and Kabaila and Giri (2009a), that may be used to motivate this restriction. For the sake of brevity,
these arguments are omitted. We also require the functions $b$ and $s$ to satisfy the
following restriction.

\smallskip

\noindent {\underbar{Restriction 2}} \ \ $b$ and $s$ are
continuous functions.

\smallskip
\noindent This implies that the endpoints of the simultaneous confidence intervals
corresponding to the confidence cube
$J(b,s)$ are continuous functions of the data. Finally, we require the
confidence cube $J(b,s)$ to coincide with the standard $1-\alpha$
confidence cube $I$ when the data strongly contradict the prior
information. The statistic $|\hat \beta_{12}|/(\hat \sigma
\sqrt{v_{55}})$ provides some indication of how far away
$\beta_{12} / (\sigma \sqrt{v_{55}})$ is from 0. We therefore
require that the functions $b$ and $s$ satisfy the following
restriction.
\smallskip

\noindent {\underbar{Restriction 3}} \ \  $b(x)=0$ for all $|x| \ge r$ and
$s(x)=d_{1-\alpha}$ for all $x \ge r$ where $r$ is a
(sufficiently large) specified positive number.

\medskip

As before, define $\gamma = \beta_{12}/\sqrt{\text{var}(\hat \beta_{12})} = \beta_{12}/(\sigma \sqrt{v_{55}})$.
As stated in Theorem 1 (see the appendix), for given functions $b$ and $s$, the coverage probability of
$J(b,s)$ is an even function of $\gamma$. We denote this coverage
probability by $c(\gamma;b,s)$.
Part of our evaluation of the confidence cube $J(b,s)$ consists of
comparing it with the standard $1-\alpha$ confidence cube $I$
using the scaled expected volume of $J(b, s)$, defined to be 
\begin{equation*}
\frac{\text{expected volume of $ J(b,
s)$}} {\text{expected volume of $I$}}.
\end{equation*}
As stated in Theorem 1 (see the appendix),
this is an even function of $\gamma$ for given function $s$. We denote this
function by $e(\gamma;s)$.
When $J(b, s)$ is a $1-\alpha$ confidence cube,
we use $\sqrt{e(\gamma;s)}$ to measure the efficiency of
the standard $1-\alpha$ confidence cube relative to
$J(b, s)$, for parameter value $\gamma$.
The reason for this is that $\sqrt{e(\gamma;s)}$ is a measure
of the ratio of sample sizes required for the expected volumes
of these two confidence cubes to be equal.

Our aim is to find functions $b$ and $s$ that satisfy Restrictions 1--3
and such that (a)
the minimum of $c(\gamma;b,s)$ over $\gamma$ is $1-\alpha$ and (b)
%
\begin{equation}
\label{criterion_simpler} \lambda \int_{-\infty}^{\infty} (e(\gamma;s) - 1) \, d
\gamma + \big ( e(0;s) - 1 \big )
\end{equation}
is minimized,
where $\lambda$ is a specified nonnegative tuning parameter.
The larger the value of $\lambda$, the smaller the relative weight
given to minimizing $e(\gamma;s)$ for $\gamma=0$, as opposed to
minimizing $e(\gamma;s)$ for other values of $\gamma$. The tuning parameter
$\lambda$ and the functions $b$ and $s$ are chosen by the statistician prior
to looking at the data.

Theorem 1, stated and proved in the appendix,
provides computationally convenient
expressions for the coverage probability of $J(b, s)$, the scaled expected
volume of $J(b, s)$ and the criterion \eqref{criterion_simpler}. The fact that this coverage probability can be
expressed as a triple integral, as opposed to a higher-dimensional
integral, is due to the symmetries of the context
considered in the present paper.

For computational feasibility, we specify the following parametric
forms for the functions $b$ and $s$. We require $b$ to be a
continuous function and so it is necessary that $b(0)=0$. Suppose
that $x_1, \ldots, x_m$ satisfy $0 = x_1 < x_2 < \cdots < x_m =
r$. Obviously, $b(x_1)=0$, $b(x_m)=0$ and
$s(x_m)=d_{1-\alpha}$. The function $b$ is fully specified by
the vector $\big (b(x_2), \ldots, b(x_{m-1}) \big)$ as follows.
Because $b$ is assumed to be an odd function, we know that
$b(-x_i) = -b(x_i)$ for $i=2,\ldots,m$. We specify the value of
$b(x)$ for any $x \in [-r,r]$ by cubic spline interpolation for
these given function values, subject to the constraint that
$b^{\prime}(-r)=0$ and $b^{\prime}(r)=0$. We fully specify the
function $s$ by the vector $\big (s(x_1), \ldots, s(x_{m-1})
\big)$ as follows. The value of $s(x)$ for any $x \in [0,r]$ is
specified by the natural cubic spline interpolation for these given function
values. We call $x_1, x_2, \ldots x_m$ the knots.

Consider the case that $c=2$ and $1-\alpha = 0.95$.
Suppose that $r = 8$, $\lambda =
0.08$ and that the knots $x_i$ at $0, r/6, \ldots,
r$ are evenly-spaced.
All computations presented in the paper were performed with programs written in MATLAB, using the optimization
and statistics toolboxes.
The resulting functions $b$ and $s$, which specify the new
0.95 confidence cube for $\theta$ are plotted in Figure 1.
The knots are denoted by small circles.
The performance of this
confidence cube is shown in Figure 2. This confidence cube has the
attractive property that its coverage probability is 0.95
throughout the parameter space. When the prior information is
correct (i.e. $\gamma = 0$), we gain since $\sqrt{e(0;s)} =
0.8558$. The maximum value of $\sqrt{e(\gamma;s)}$ is $1.0956$.
This confidence cube coincides with the standard 0.95
confidence cube for $\theta$ when the data strongly contradicts the prior
information, so that $\sqrt{e(\gamma;s)}$ approaches 1 as $\gamma
\rightarrow \infty$.

\FloatBarrier
\begin{figure}[h]
\label{Figure1} \hspace{-2.3cm}
\includegraphics[scale=0.95]{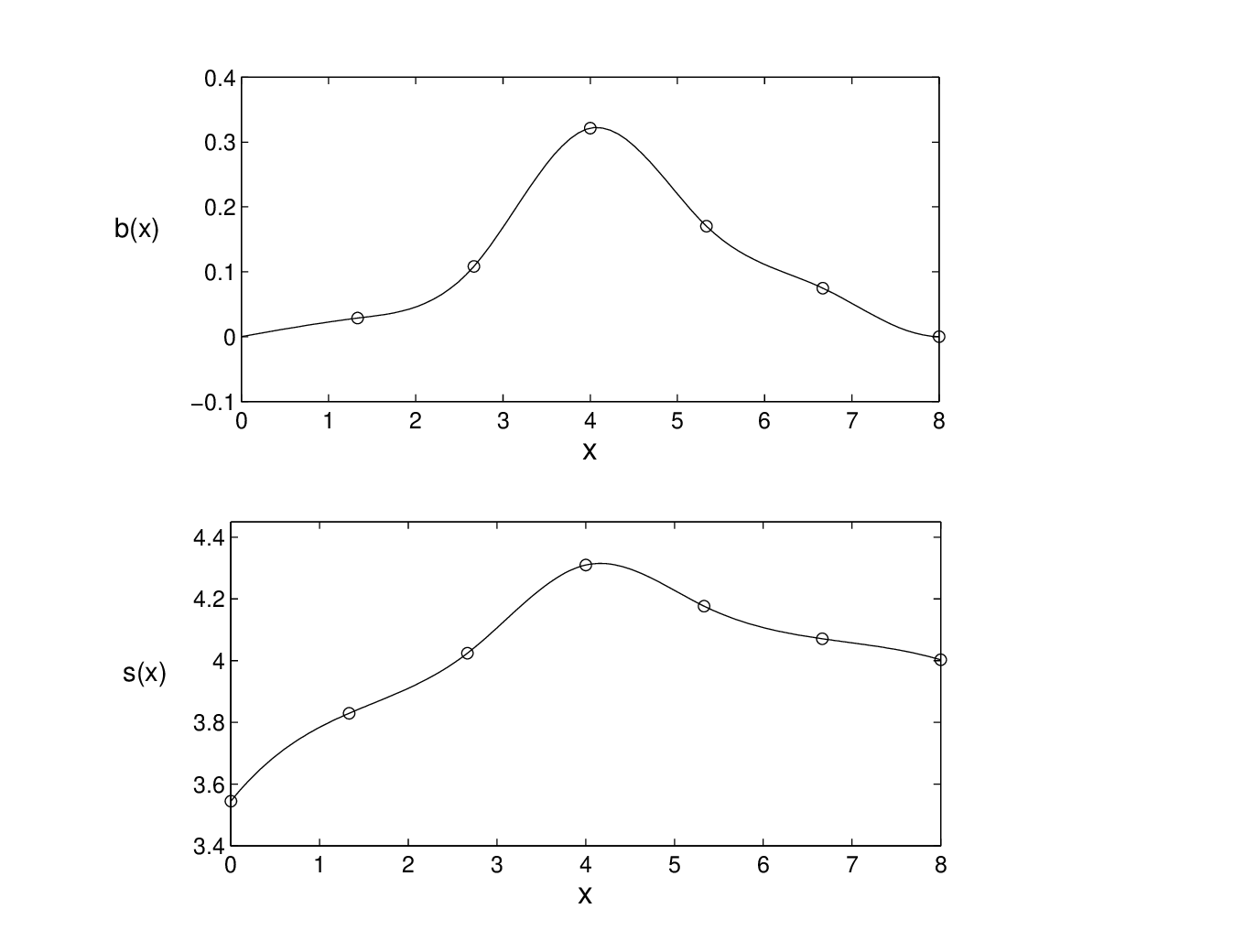}
       \caption{ Plots of the functions $b$ and $s$ that specify the new 0.95 confidence cube for
       $\theta = (\theta_{00}, \theta_{10}, \theta_{01},
\theta_{12})$, when $c=2$, $r=8$ and
       $\lambda=0.08$.}
\end{figure}
\FloatBarrier

\FloatBarrier
\begin{figure}[h]
\label{Figure2}\hspace{-1.3cm}
\includegraphics[scale=0.95]{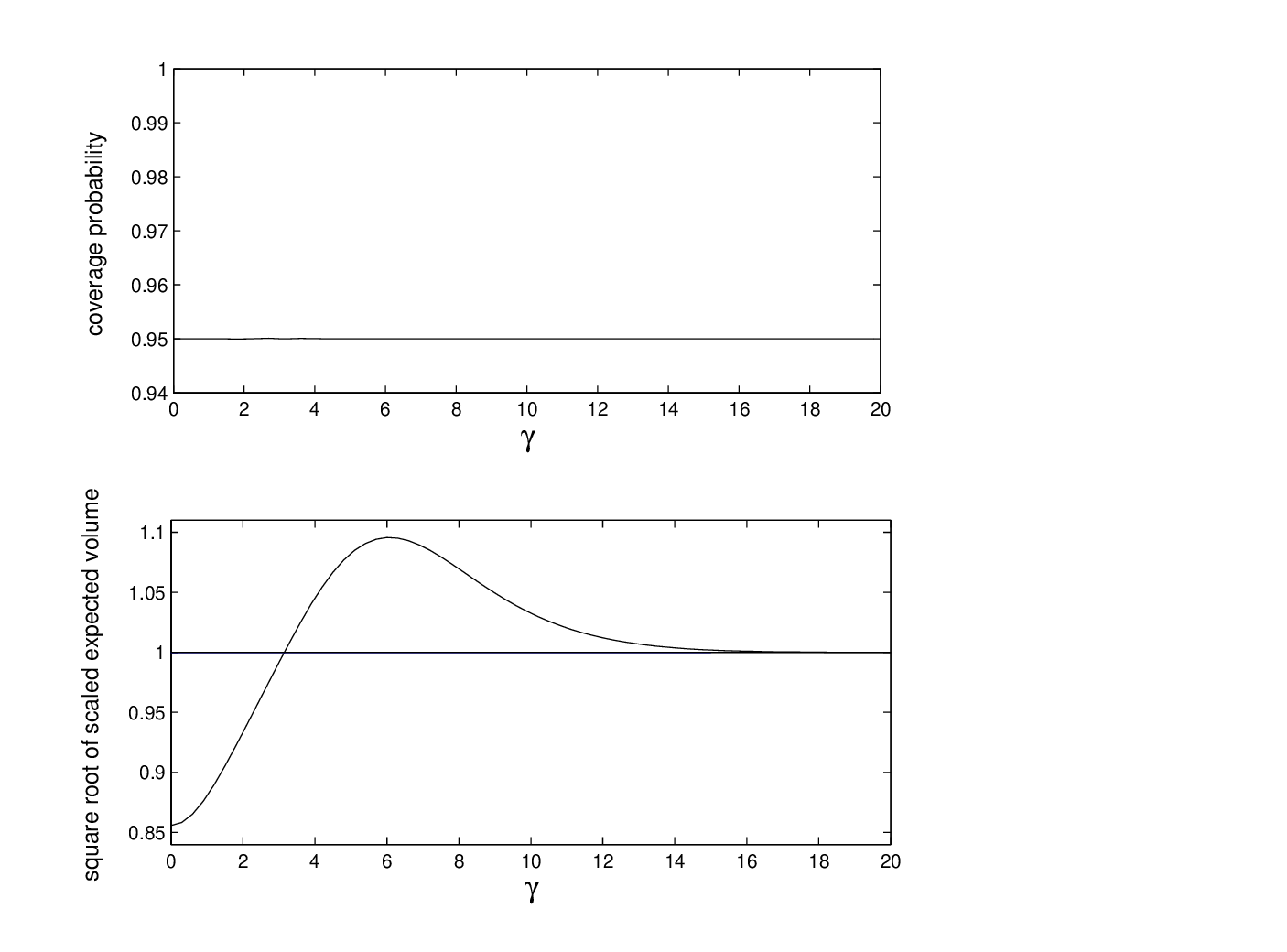}
       \caption{ Plots of the coverage probability and the square root of the scaled expected volume $e(\gamma;s)$
{\big(}as functions of $\gamma = \beta_{12}/\sqrt{\text{var}(\hat
\beta_{12})}${\big )} of the new 0.95 confidence cube for
$\theta = (\theta_{00}, \theta_{10}, \theta_{01}, \theta_{12})$. This cube
was obtained using $c=2$, $r=8$ and
       $\lambda=0.08$.}
\end{figure}
\FloatBarrier

\baselineskip=21pt

\bigskip

\noindent {\bf 4. Illustration of the Application of the New Confidence Cube}

\medskip

\noindent In this section we illustrate the application of the new $1-\alpha$ confidence cube,
utilizing the uncertain prior information, to a real data set.
We extract a $2 \times 2$ factorial data set from the $2^3$ factorial data set described in Table 7.5
of Box et al (1963) as follows. Define $x_1 = -1$ and $x_1 = 1$ for ``Time of addition of HNO$_3$'' equal
to 2 hours and 7 hours, respectively.  Also define $x_2 = -1$ and $x_2 = 1$ for ``heel absent'' and
``heel present'', respectively. The observed responses (percent yield of a nitration process) are the following: \newline
\phantom{1}\ \ \ \ \ \ For $(x_1,x_2) = (-1,-1)$, $y=87.2$. \newline
\phantom{1}\ \ \ \ \ \ For $(x_1,x_2) = (1,-1)$, $y=88.4$. \newline
\phantom{1}\ \ \ \ \ \ For $(x_1,x_2) = (-1,1)$, $y=86.7$. \newline
\phantom{1}\ \ \ \ \ \ For $(x_1,x_2) = (1,1)$, $y=89.2$. \newline
We use the model \eqref{regression}. The discussion on p.265 of Box et al (1963) implies that there is
uncertain prior information that $\beta_{12} = 0$. The discussion on p.266 of Box et al (1963) implies that there is
an estimator $\hat \sigma^2$ of $\sigma^2$, obtained from other related experiments, with the property
that $\hat \sigma^2/\sigma^2 \sim Q/m$ where $Q \sim \chi^2_m$ and $m$ is effectively infinite.
The observed value of $\hat \sigma$ is 0.8.
The standard 0.95 confidence cube for $\theta$ is
\begin{equation*}
[87.2 \pm 1.99272] \times [88.4 \pm 1.99272] \times [86.7 \pm 1.99272] \times [89.2 \pm 1.99272].
\end{equation*}
We have also computed the new 0.95 confidence cube for $\theta$, using $d=6$, $\lambda=0.08$
and equally-spaced knots at $0,1,\ldots,6$. This new confidence cube is
\begin{equation*}
[87.1748 \pm 1.88504] \times [88.4252 \pm 1.88504] \times [86.7252 \pm 1.88504] \times [89.1748 \pm 1.88504].
\end{equation*}
Thus
\begin{equation*}
\sqrt{\frac{(\text{volume of new 0.95 confidence cube})}
{(\text{volume of standard 0.95 confidence cube})}} = 0.8948
\end{equation*}
For this data set, we have clearly gained by using the new 0.95 confidence cube for $\theta$.


\newpage

\bigskip

\noindent {\bf {5. Remarks}}

\medskip

\noindent {\underbar{\sl Remark 5.1}} \ The criterion \eqref{criterion_simpler}
can be expresses as the weighted average
\begin{equation*}
\label{criterion} \int_{-\infty}^{\infty} (e(\gamma;s) - 1) \, d
\nu(\gamma),
\end{equation*}
where the weight function $\nu$ has been chosen to
be
\begin{equation*}
\label{mixed_wt_fn} \nu(x) = \lambda x + {\cal H}(x) \ \text{ for
all } \ x \in \mathbb{R},
\end{equation*}
where $\lambda$ is a specified nonnegative number and ${\cal H}$
is the unit step function defined by ${\cal H}(x) = 0$ for $x < 0$
and ${\cal H}(x) = 1$ for $x \ge 0$.
The idea of minimizing a weighted average expected length of a confidence
interval, subject to a coverage probability constraint, is due to Pratt (1961).
The particular weight function $\nu$ was first used in
related contexts by Farchione and Kabaila (2008) and Kabaila and Giri (2009c).

\medskip

\noindent {\underbar{\sl Remark 5.2}} \ An attempt to utilize the uncertain prior information is as
follows. We carry out a preliminary test of the null hypothesis
that the two-factor interaction is zero against the alternative
hypothesis that it is non-zero. If this null hypothesis is
accepted then the confidence cube for
$\theta$, with nominal
confidence coefficient $1-\alpha$, is constructed
assuming that it is known {\sl a priori} that the two-factor
interaction is zero; otherwise the standard
$1-\alpha$ confidence cube is used. We call this the naive
$1-\alpha$ confidence cube for
$\theta$. This assumption is false
and it leads to a naive $1-\alpha$ confidence cube with
minimum coverage probability less than $1-\alpha$. For
example, for $\alpha = 0.05$, $c=2$ and a preliminary test with level of
significance 0.05, this minimum coverage probability is
0.9078 (Giri, 2008). The poor coverage properties of these naive confidence
cubes are presaged by the following two strands of literature.
The first strand concerns the poor properties of inferences about
main effects after preliminary hypothesis tests in factorial
experiments, see e.g. Neyman (1935), Traxler (1976), Bohrer and
Sheft (1979), Fabian (1991), Shaffer (1991) and Ng (1994). The
second strand concerns the poor coverage properties of naive
(non-simultaneous) confidence intervals in the context of linear
regression models with zero-mean normal errors, see e.g. Kabaila (1998, 2005, 2009),
Kabaila and Leeb (2006), Giri and Kabaila (2008) and
Kabaila and Giri (2009c).

\medskip

\noindent {\underbar{\sl Remark 5.3}} \
Whilst the naive $1-\alpha$ confidence cube for $\theta$ (defined in Remark 5.2)
fails to properly utilize the prior information,
its form can be used to provide some motivation for the new
$1-\alpha$ confidence cube for $\theta$, described in Section 3, that utilizes the uncertain prior information.

The naive $1-\alpha$ confidence cube for $\theta$ is obtained as follows.
The usual test statistic for testing the null hypothesis that
$\beta_{12} = 0$ against the alternative hypothesis that
$\beta_{12} \ne 0$ is $\hat \beta_{12}/(\hat \sigma
\sqrt{v_{55}})$. This test statistic has a $t_{n-p}$ distribution
under this null hypothesis. Suppose that, for some given positive
number $q$, we fix $\beta_{12}$ at 0 if $|\hat \beta_{12}|/(\hat
\sigma \sqrt{v_{55}}) \le q$; otherwise we allow $\beta_{12}$ to
vary freely.
Define
$\tilde \sigma^2 = \big ( (n-p)\hat \sigma^2 + (\hat \beta_{12}^2/v_{55})\big)/(n-p+1)$.
This is the usual estimator of $\sigma^2$ obtained by fitting the full model to
the data when it is assumed that $\beta_{12} = 0$. Define
$\tilde W = \tilde \sigma/ \sigma$. Note that $\tilde W$ has the same distribution
as $\sqrt{\tilde Q/(n-p+1)}$, where $\tilde Q \sim \chi^2_{n-p+1}$ when $\beta_{12} = 0$.
Suppose that $Z_1$, $Z_2$, $Z_3$ and
$\tilde W$ are independent random variables and $Z_i \sim N(0,1)$ ($i=1,2,3$).
Let $\tilde Z_1 = (Z_1 - Z_2 - Z_3)/\sqrt{3}$,
$\tilde Z_2 = (Z_1 + Z_2 - Z_3)/\sqrt{3}$,
$\tilde Z_3 = (Z_1 - Z_2 + Z_3)/\sqrt{3}$
and $\tilde Z_4 = (Z_1 + Z_2 + Z_3)/\sqrt{3}$.
Define $\tilde d_{1-\alpha}$ by
$P \big ( \max_{i=1,\ldots,4} |\tilde Z_i|/\tilde W \le \tilde d_{1-\alpha} \big) = 1-\alpha$.
The
naive $1-\alpha$ confidence cube is obtained as follows. If $|\hat
\beta_{12}|/(\hat \sigma \sqrt{v_{55}}) > q$ then this confidence
cube is $I$.
If, on the other hand, $|\hat \beta_{12}|/(\hat \sigma \sqrt{v_{55}}) \le q$ then this
confidence cube is
\begin{align*}
&\big[ \hat \Theta_{00} - \hat \beta_{12} \pm (\sqrt{3/c}/2) \tilde d_{1-\alpha} \tilde \sigma \big]
\times
\big[ \hat \Theta_{10} + \hat \beta_{12} \pm (\sqrt{3/c}/2) \tilde d_{1-\alpha} \tilde \sigma \big]\\
&\phantom{1234} \times\big[ \hat \Theta_{01} + \hat \beta_{12} \pm (\sqrt{3/c}/2) \tilde d_{1-\alpha} \tilde \sigma \big]
\times
\big[ \hat \Theta_{110} - \hat \beta_{12} \pm (\sqrt{3/c}/2) \tilde d_{1-\alpha} \tilde \sigma \big]
\end{align*}
The naive $1-\alpha$ confidence cube can be expressed in the form \eqref{J(b,s)}
where
\begin{align*}
&b(x) =  \begin{cases}
        0 &\ \ \ \ \ \text{for } \ \ \ |x| > q\\
        x/2 &\ \ \ \ \ \text{for} \ \ \ \ |x| \le q.
    \end{cases} \\
&s(x) =  \begin{cases}
        d_{1-\alpha}  &\text{for } \ \ \ x > q\\
        \tilde d_{1-\alpha} (\sqrt{3}/2) \sqrt{(n-p + x^2)/(n-p+1)}&\text{for} \ \ \ \ 0 < x \le q.
    \end{cases}
\end{align*}
The fact that the naive confidence cube can be expressed in terms of just
two functions $b$ and $s$ is a reflection of the symmetries of the context
considered in the present paper. The confidence cube \eqref{J(b,s)}, with $b$ and $s$ satisfying
only Requirements 1--3,
is similar in form to the naive
$1-\alpha$ confidence cube, but
with a great ``loosening up'' of the forms that the functions $b$
and $s$ can take.

\medskip

\noindent {\underbar{\sl Remark 5.4}} \ The new $1-\alpha$ confidence cube is computed to satisfy the
constraint that its minimum coverage probability is $1-\alpha$.
For the examples described in both
the previous section (for which $c=2$, $r=8$ and $\lambda=0.08$)
and the current section (for which $c$ is effectively infinite, $r=6$ and $\lambda=0.08$), it is remarkable
that the new 0.95 confidence cube has coverage probability {\sl equal} to 0.95
throughout the parameter space. The new 0.95 confidence cube was also computed for (a) $c=3$, $r=8$ and $\lambda=0.08$,
(b) $c=7$, $r=8$ and $\lambda=0.08$ and (c) $c=20$, $r=8$ and $\lambda=0.08$. In each of these cases,
the new 0.95 confidence cube also has coverage probability {\sl equal} to 0.95
throughout the parameter space. This provides strong empirical evidence that the new $1-\alpha$ confidence cube
has the attractive property that its coverage probability is equal to $1-\alpha$ throughout the parameter space.

\bigskip

\noindent {\bf {Appendix: Theorem 1}}

\medskip

\noindent In this appendix, we provide computationally convenient expressions for
the coverage probability of $J(b, s)$, the scaled expected
volume of $J(b, s)$ and the criterion \eqref{criterion_simpler}.
Let $f_W$ denote the probability density
function of $W$.

\medskip

\noindent {\bf Theorem 1.}

\noindent (a) {\sl Define $\ell_1 = 2 \ell(h,w) + t_3 - h + \gamma$,
$u_1 = 2 u(h,w) + t_3 - h + \gamma$,
$\ell_2 = -2 u(h,w) + t_3 + h - \gamma$ and
$u_2 = -2 \ell(h,w) + t_3 + h - \gamma$. Now define
$\tilde \ell_1 = \max(\ell_1, -u_1)$,
$\tilde u_1 = \min(u_1, - \ell_1)$,
$\tilde \ell_2 = \max(\ell_2, -u_2)$ and
$\tilde u_2 = \min(u_2, - \ell_2)$. We use these
functions to define
\begin{equation*}
k(t_{3},h,w,\gamma) =
\begin{cases}
0 \text{ if either } \tilde \ell_1>\tilde u_1 \text{ or }\tilde \ell_2>\tilde u_2, \\
\Big(\Phi \big(\tilde u_1/\sqrt{2} \big) - \Phi \big(\tilde
\ell_1/\sqrt{2} \big)\Big) \left(\Phi \big(\tilde u_2/\sqrt{2} \big)
-\Phi \big(\tilde \ell_2/\sqrt{2} \big)\right)
  \text{ otherwise,}
\end{cases}
\end{equation*}
where $\Phi$ denotes the $N(0,1)$ distribution function.

\smallskip

Define $\ell^{\dag}(w) = - d_{1-\alpha} w$ and
$u^{\dag}(w) = d_{1-\alpha} w$.
Also define $\ell_1^{\dag} = 2 \ell^{\dag}(w) + t_3 - h + \gamma$,
$u_1^{\dag} = 2 u^{\dag}(w) + t_3 - h + \gamma$,
$\ell_2^{\dag} = -2 u^{\dag}(w) + t_3 + h - \gamma$ and
$u_2^{\dag} = -2 \ell^{\dag}(w) + t_3 + h - \gamma$. Now define
$\tilde \ell_1^{\dag} = \max(\ell_1^{\dag}, -u_1^{\dag})$,
$\tilde u_1^{\dag} = \min(u_1^{\dag}, - \ell_1^{\dag})$,
$\tilde \ell_2^{\dag} = \max(\ell_2^{\dag}, -u_2^{\dag})$ and
$\tilde u_2^{\dag} = \min(u_2^{\dag}, - \ell_2^{\dag})$. We use these
functions to define
\begin{equation*}
k^{\dag}(t_{3},h,w,\gamma) =
\begin{cases}
0 \text{ if either } \tilde \ell_1^{\dag}>\tilde u_1^{\dag} \text{ or }\tilde \ell_2^{\dag}>\tilde u_2^{\dag}, \\
\Big(\Phi \big(\tilde u_1^{\dag}/\sqrt{2} \big) - \Phi \big(\tilde
\ell_1^{\dag}/\sqrt{2} \big)\Big) \left(\Phi \big(\tilde u_2^{\dag}/\sqrt{2} \big)
-\Phi \big(\tilde \ell_2^{\dag}/\sqrt{2} \big)\right)
  \text{ otherwise.}
\end{cases}
\end{equation*}

The coverage probability $P \big(\theta \in J(b, s) \big)$
is equal to
\begin{equation}
\label{cov_prob} (1-\alpha) + \int_0^{\infty} \int_{-r}^{r}
\int_{-\infty}^{\infty} \big( k(t_{3},wx,w,\gamma) -
k^{\dag}(t_3,wx,w, \gamma) \big) \,\phi(t_3)\,dt_3\,
\phi(wx-\gamma)\, dx \, w \, f_W(w) \, dw
\end{equation}
where $\phi$ denotes the $N(0,1)$ probability density function.
For given functions $b$ and $s$, this is an even function of
$\gamma$.}

\smallskip

\noindent (b) {\sl The scaled expected volume of $J(b, s)$ is
equal to
\begin{equation}
\label{comp_conv_e}
1 + \frac{1} {d_{1-\alpha}^4 \, E(W^4)}
 \int^{\infty}_0 \int^{r}_{- r} \left (s^4(|x|) - d^4_{1-\alpha} \right )
\phi(w x -\gamma) \, dx \,  w^5 \,  f_W(w)  \, dw.
\end{equation}
For given function $s$, this is an even function of $\gamma$.
}

\smallskip

\noindent (c) {\sl The criterion \eqref{criterion_simpler} is equal to
\begin{equation}
\label{criterion_final}
\frac{2} {d_{1-\alpha}^4 \, E(W^4)}
 \int^{\infty}_0 \int^{r}_{0} \left (s^4(x) - d^4_{1-\alpha} \right )
(\lambda + \phi(w x)) \, dx \,  w^5 \,  f_W(w)  \, dw.
\end{equation}
}

\smallskip

\noindent {\bf Proof of part (a).}

\smallskip

\noindent
Define
$T_{1}=(\hat
\beta_{0}-\beta_{0})/(\sigma \sqrt{v_{55}})$, $T_{2}=(\hat
\beta_{1}-\beta_{1})/(\sigma \sqrt{v_{55}})$, $T_{3}=(\hat
\beta_{2}-\beta_{2})/(\sigma \sqrt{v_{55}})$ and  $H = \hat
\beta_{12}/(\sigma \sqrt{v_{55}})$.
Then define $G_{1} = (T_{1}-T_{2}-T_3+H-\gamma)/2$,
$G_{2}=(T_{1}+T_{2}-T_3-H+\gamma)/2$,
$G_{3}=(T_{1}-T_{2}+T_3-H+\gamma)/2$,
$G_{4}=(T_{1}+T_{2}+T_3+H-\gamma)/2$. Note that $H \sim N(\gamma,1)$.
It is straightforward to show that the coverage probability
$P\big( \theta \in
J(b, s) \big)$ is equal to
\begin{multline}
\label{cov_pr_J(b,s)}
P \big ( \ell(H,W) \le G_{1} \le u(H,W), \, \ell(H,W) \le -G_{2} \le
u(H,W), \\
\ell(H,W) \le -G_{3} \le u(H,W), \, \ell(H,W) \le G_{4} \le
u(H,W) \big )
\end{multline}
where the functions $\ell(\cdot, \cdot) : \mathbb{R} \times [0,
\infty) \rightarrow \mathbb{R}$ and $u(\cdot, \cdot) : \mathbb{R}
\times [0, \infty) \rightarrow \mathbb{R}$  are defined by
$\ell(h,w) = b(h/w)w - s(|h|/w)w$ and
$u(h,w) =  b(h/w)w + s(|h|/w)w$.

 It follows from
the $N(\gamma,1)$ distribution of $H$  and the
independence of the random vectors $(G_{1}, G_{2}, G_{3}, G_{4}, H)$ and $W$ that
\eqref{cov_pr_J(b,s)} is equal to
\begin{equation*}
\int_0^{\infty} \int_{-\infty}^{\infty} a(h,w) \, \phi(h-\gamma) \, dh \, f_W(w) \, dw,
\end{equation*}
where
\begin{align*}
a(h,w) = &P \big (\ell(h,w) \le G_1 \le u(h,w), \, \ell(h,w) \le -G_2 \le u(h,w),\\
&\phantom{1234567890}\ell(h,w) \le -G_3 \le u(h,w), \, \ell(h,w) \le G_4 \le u(h,w)\, \big | \, H=h \big).
\end{align*}
Note that $T_1$, $T_2$, $T_3$ and $H$ are independent random variables and that
$T_1$, $T_2$ and $T_3$ are identically
$N(0,1)$ distributed. Thus $T_1 - T_2$ and $T_1 + T_2$ are independent $N(0,2)$ distributed random variables
and $a(h,w)$ is equal to
\begin{align*}
&\int_{-\infty}^{\infty} P \big ( \ell_1 \le T_1 - T_2 \le u_1, \, -u_1 \le T_1 - T_2 \le -\ell_1, \\
&\phantom{12345678901234567890123456}\ell_2 \le T_1 + T_2 \le u_2, \, -u_2 \le T_1 + T_2 \le -\ell_2  \big)
\, \phi(t_3) \, dt_3 \\
&= \int_{-\infty}^{\infty} P \big ( \tilde \ell_1 \le T_1 - T_2 \le \tilde u_1)
P \big ( \tilde \ell_2 \le T_1 + T_2 \le \tilde u_2) \, \phi(t_3) \, dt_3 \\
&= \int_{-\infty}^{\infty} k(t_3, h, w, \gamma) \, \phi(t_3) \, dt_3
\end{align*}
Thus $P(\theta \in J(b,s))$ is equal to
\begin{equation}
\int_0^{\infty} \int_{-\infty}^{\infty} \int_{-\infty}^{\infty} k(t_3, h, w, \gamma) \, \phi(t_3) \, dt_3
\, \phi(h-\gamma) \, dh \, f_W(w) \, dw.
\label{cov}
\end{equation}
The standard $1-\alpha$ confidence cube $I$ has coverage
probability $1-\alpha$. Hence
\begin{equation}
\label{cov_stand}
1-\alpha = \int_0^{\infty}
\int_{-\infty}^{\infty} \int_{-\infty}^{\infty}\,
k^\dag(t_{3},h,w,\gamma)\, \phi(t_3)\,dt_3\, \phi(h-\gamma)\, dh
\, f_W(w) \, dw.
\end{equation}
Subtracting \eqref{cov_stand} from \eqref{cov}  and noting that,
by Restriction 3, $b(x)=0$ for all $|x| \ge r$
and $s(x)=d_{1-\alpha}$ for all $x \ge r$, we find that $P(\theta \in J(b,s))$
is equal to
\begin{equation*}
(1-\alpha) +
\int_0^{\infty} \int_{-rw}^{rw}\int_{-\infty}^{\infty} \big(
k(t_3,h,w, \gamma) - k^{\dag}(t_3,h,w\gamma) \big)
\,\phi(t_3)\,dt_3\, \phi(h-\gamma)\,  dh \, f_W(w) \, dw.
\end{equation*}
Changing the variable of integration from $h$ to $x=h/w$, we obtain \eqref{cov_prob}.
It is straightforward to show that $P(\theta \in J(b,s))$ is an even function of
$\gamma$.

\smallskip

\noindent{\bf Proof of part (b).}

\smallskip

\noindent The scaled expected volume of $J(b,s)$ is
equal to
\begin{equation*}
\frac{E \left (W^4 \, s^4 \bigg( \displaystyle{\frac{|H|}{W}}\bigg)
\right )} {d_{1-\alpha}^4 \, E(W^{4})}.
\end{equation*}
For given function $s$, this is an even function of $\gamma$. We denote this function by
$e(\gamma;s)$.
The random variables $H$ and $W$ are independent.  Thus
\begin{equation}
\label{sc_exp_len}
 e(\gamma;s) =
 \frac{1} {d^4_{1 - \alpha} \, E(W^4)}
 \int^{\infty}_0 \int^{\infty}_{-\infty} s^4 \left (\frac{|h|}{w}
\right) \phi(h-\gamma) \, dh \,  w^4 \,  f_W(w)  \, dw
\end{equation}
Obviously,
\begin{equation}
\label{obvious}
 1 = \frac{1} {d^4_{1 - \alpha} \, E(W^4)}
 \int^{\infty}_0 \int^{\infty}_{-\infty} d^4_{1 - \alpha} \, \phi(h-\gamma)
 \, dh \,  w^4 \,  f_W(w)  \, dw.
\end{equation}
Note that $s(x)=d_{1-\alpha}$ for all $x \ge r$. Subtracting
\eqref{obvious} from \eqref{sc_exp_len}, we therefore obtain
\begin{equation*}
 e(\gamma;s) = 1 +
 \frac{1} {d^4_{n-p, 1 - \alpha} \, E(W^4)}
 \int^{\infty}_0 \int^{r w}_{- r w} \left (s^4 \left (\frac{|h|}{w}
\right) - d^4_{1-\alpha} \right ) \phi(h-\gamma) \, dh \, w^4
\, f_W(w)  \, dw.
\end{equation*}
Changing the variable of integration from
$h$ to $x = h/w$, we obtain \eqref{comp_conv_e}.

\smallskip

\noindent{\bf Proof of part (c).}

\smallskip

\noindent Substituting \eqref{comp_conv_e} into \eqref{criterion_simpler}, we obtain
that \eqref{criterion_simpler} is equal to
\begin{equation*}
\frac{1} {d^4_{1-\alpha} \, E(W^4)}
 \int^{\infty}_0 \int^{r}_{- r} \left (s^4(|x|) - d^4_{1-\alpha} \right )
\left ( \lambda \, \int_{-\infty}^{\infty} \phi(w x -\gamma) \, d \gamma + \phi(wx) \right ) \, dx \,  w^5 \,  f_W(w)  \, dw,
\end{equation*}
which is equal to \eqref{criterion_final}.

\bigskip

\noindent {\bf References}

\medskip

\rf Bickel, P.J. (1984). Parametric robustness: small biases can be
worthwhile. {\sl Annals of Statistics} 12:864--879.

\rf Bickel, P.J., Doksum, K.A. (1977).  {\sl Mathematical Statistics,
Basic Ideas and Selected Topics}. Oakland, California: Holden-Day.

\rf Bohrer,  R., Sheft, J. (1979). Misclassification probabilities
in $2^3$ factorial experiments. {\sl Journal of Statistical Planning
and Inference} 3:79--84.

\rf Box, G.E.P., Connor, L.R., Cousins, W.R., Davies, O.L.,
Hinsworth, F.R. \& Sillitto, G.P (1963). {\sl The Design and Analysis
of Industrial Experiments}, 2nd edition, reprinted,
London: Oliver and Boyd.

\rf Brown, L.D., Casella, G., HWANG, J.T.G. (1995). Optimal confidence sets, bioequivalence and the Limacon of Pascal.
{\sl Journal of the American Statistical Association} 90:880--889.

\rf Fabian, V. (1991). On the problem of interactions in the
analysis of variance. {\sl Journal of the American Statistical
Association} 86:362--373.

\rf Giri, K. (2008). Confidence intervals in regression utilizing prior information.
Unpublished PhD thesis, Department of Mathematics and Statistics, La Trobe University.

\rf Giri, K., Kabaila, P. (2008).  The coverage probability of
confidence intervals in $2^r$ factorial experiments after
preliminary hypothesis testing. {\sl Australian \& New
Zealand Journal of Statistics} 50:69--79.

\rf Farchione, D., Kabaila, P. (2008). Confidence intervals for the
normal mean utilizing prior information. {\sl Statistics
and Probability Letters} 78:1094--1100.

\rf Hodges, J.L., Lehmann, E.L. (1952). The use of previous
experience in reaching statistical decisions. {\sl Annals of
Mathematical Statistics} 23:396--407.

\rf Kabaila, P. (1998). Valid confidence intervals in regression
after variable selection. {\sl Econometric Theory} 14:463--482.

\rf Kabaila, P. (2005). On the coverage probability of confidence
intervals in regression after variable selection. {\sl Australian \&
New Zealand Journal of Statistics} 47:549--562.

\rf Kabaila P. (2009). The coverage properties of confidence regions after model
selection. {\sl International Statistical Review} 77:405--414.

\rf Kabaila, P. \&  Giri, K. (2009a). Confidence intervals in regression utilizing prior
information. {\sl Journal of Statistical Planning and Inference}
139:3419--3429.

\rf Kabaila, P., Giri, K. (2009b). Large-sample confidence intervals
for the treatment difference in a two-period crossover trial, utilizing
prior information.
{\sl Statistics and Probability Letters} 79:652--658.


\rf Kabaila, P.,  Giri, K. (2009c). Upper bounds on the minimum
coverage probability of confidence intervals in regression after
variable selection. 
{\sl Australian \& New Zealand Journal of Statistics} 51:271--288.

\rf Kabaila, P., Leeb, H. (2006). On the large-sample minimal
coverage probability of confidence intervals after model
selection. {\sl Journal of the American Statistical Association} 101:619--629.

\rf Kabaila, P., Tuck, J. (2008). Confidence intervals utilizing
prior information in the Behrens-Fisher problem.
{\sl Australian \& New Zealand Journal of Statistics} 50:309--328.

\rf Maatta, J.M. \& Casella, G. (1990). Decision-theoretic estimation. {\sl Statistical
Science} 5:90--120.

\rf Kempthorne, P.J. (1983). Minimax-Bayes compromise estimators. In
{\sl 1983 Business and Economic Statistics Proceedings of the
American Statistical Association}, Washington DC, pp.568--573.


\newpage

\rf Kempthorne, P.J. (1987).  Numerical specification of
discrete least favourable prior distributions.  {\sl SIAM
Journal on Scientific and Statistical Computing} 8:171--184.


\rf Kempthorne, P.J. (1988). Controlling risks under different loss
functions: the compromise decision problem. {\sl Annals of
Statistics} 16:1594--1608.

\rf Maatta, J.M., Casella, G. (1990). Decision-theoretic estimation. {\sl Statistical
Science} 5:90--120.

\rf Neyman, J. (1935). Comments on `Complex experiments' by F.
Yates. {\sl Supplement to the Journal of the Royal Statistical Society}
2:181--247.

\rf  Ng, T-H. (1994). The impact of a preliminary test for
interaction in a $2 \times 2$ factorial trial. {\sl Communications in
Statistics: Theory and Methods} 23:435--450.

\rf Pratt, J.W. (1961). Length of confidence intervals. {\sl Journal of
the American Statistical Association} 56:549--657.

\rf Puza, B., O'Neill, T. (2006a). Generalised Clopper-Pearson confidence intervals for the binomial proportion.
{\sl Journal of Statistical Computation and Simulation} 76:489 -- 508.

\rf Puza, B., O'Neill, T. (2006b). Interval estimation via tail functions.
{\sl Canadian Journal of Statistics} 34:299 -- 310.

\rf Saleh, A.K.Md.E. (2006). {\sl Theory of Preliminary Test and Stein-Type Estimation and Applications.}
Hoboken, NJ: Wiley.

\rf Shaffer, J.P. (1991). Probability of directional errors with
disordinal (qualitative) interaction. {\sl Psychometrika} 56:29--38.

\rf Traxler, R.H. (1976). A snag in the history of factorial
experiments, in: Owen, D.B. (Ed.), {\sl On the History of Statistics and
Probability}, Marcel Dekker, New York, pp. 281--295.

\end{document}